\documentclass[runningheads,a4paper]{llncs}
\usepackage{amssymb}
\usepackage{graphicx}
\usepackage{url}
\usepackage{xcolor}
\newcommand{\keywords}[1]{\par\addvspace\baselineskip
\noindent\keywordname\enspace\ignorespaces#1}
\usepackage{bm}
\usepackage{textcomp}

\begin{document}

\mainmatter
%================================================

%==== FILL IN ====================================
\title{TheoryGuru: A Mathematica Package to apply Quantifier Elimination Technology to Economics}  % Full title
\titlerunning{TheoryGuru: A Mathematica Package to apply QE Technology to Economics} % Short title
\author{Casey B. Mulligan\inst{1} \and  James H. Davenport \inst{2} \and Matthew England \inst{3}}
\authorrunning{Mulligan-Davenport-England}
\institute{
University of Chicago, USA \\%$-$
\email{c-mulligan@uchicago.edu}\\ 
%\texttt{url}
\and
University of Bath, UK \\%$-$
\email{J.H.Davenport@bath.ac.uk}\\ 
%\texttt{url}
\and
Coventry University, UK \\%$-$
\email{Matthew.England@coventry.ac.uk}\\ 
%\texttt{url}
}
\maketitle

\begin{abstract}
%This should be a short summary of the contents of the main body.  
%\begin{itemize}
%\item It must be self-contained (without any external references, etc)
%\item It must be in plain text (without using any mathematical symbols, etc).
%\end{itemize}
%
We consider the use of Quantifier Elimination (QE) technology for automated reasoning in economics.   
There is a great body of work considering QE applications in science and engineering but we demonstrate here that it also has use in the social sciences.  We explain how many suggested theorems in economics could either be proven, or even have their hypotheses shown to be inconsistent, automatically via QE.

However, economists who this technology could benefit are usually unfamiliar with QE, and the use of mathematical software generally.  This motivated the development of a \textsc{Mathematica} Package \texttt{TheoryGuru}, whose purpose is to lower the costs of applying QE to economics.  We describe the package's functionality and give examples of its use.
\keywords{Quantifier Elimination, Economics Reasoning}
\end{abstract}

%%%%%%%%%%%%%%%%%%%%%%%%%%%%%%%%%%%%%%%%%%%%%%%%%
%%%%--- Remove this after reading it.
%\framebox{
%\begin{minipage}{5in}
%\textbf{Guidance from ICMS:}
%The main body should describe challenge, achievements and progress in
%mathematical software research, addressing issues such as
%functionality, underlying theories, design, development and applications. 
%\begin{itemize}
% \item The whole paper (including the title page and the references) must be
%       \begin{itemize}
%       \item[] {\bf at least 4 pages} 
%       \item[] {\bf at most 8 pages}.
%       \end{itemize}
% \item For a new software, some comparison with  existing software (if exists) will be appropriate. 
% \item Carefully divide the main body into several meaningful sections. 
%\end{itemize}
%\end{minipage}
%}
%%%%%%%%%%%%%%%%%%%%%%%%%%%%%%%%%%%%%%%%%%%%%%%%%

%------------------------------------------------------------
\section{Introduction}

A general task in economic reasoning is to determine whether, with variables $\bm{v}=(v_1,\ldots,v_n)$, a hypothesis $H(\bm{v})$ follows from  assumptions $A(\bm{v})$, i.e. is it the case that
$
\forall \bm{v} \, . \,  A \Rightarrow H
$?
Ideally the answer would be \verb+True+ or \verb+False+, but in practice life is more complicated: the answer could differ depending on the value of $\bm{v}$; or the assumptions could even be contradictory, i.e. $A(v)$ alone is \verb+False+.  

We can categorise these possibilities via the outcome of a pair of quantified statements (Table \ref{tab:main}).
Should technology provide any one automatically then an economist gains important information: either a proof or a disproof of their theory; or an identification of where their theory may be true (a description of $\{\bm{v}:A(\bm{v})\Rightarrow H(\bm{v})\}$); or the knowledge that their assumptions contradict.  

\begin{table}[t]
\caption{Table of possible outcomes from a potential theorem $\forall \bm{v} \, . \, A \Rightarrow H$ \label{tab:main} }
\centering
\begin{tabular}{r|cc}
& $\lnot\exists \bm{v}[A \land \lnot H]$ & $\exists \bm{v}[A \land \lnot H]$ \\[0.1cm] \hline 
$\exists \bm{v}[A \land H]$      \,& \verb+True+          & \verb+Mixed+ \\
$\lnot\exists \bm{v}[A \land H]$ \, & \, Contradictory Assumptions & \verb+False+
\end{tabular} 
\end{table}

Such technology could also allow for exploration.  An economist could vary the question: the assumptions generating a \verb+True+ result can be weakened, or those generating a \verb+Mixed+ result strengthened, by quantifying more or less of $\bm{v}$.  

For example, we can partition $\bm{v}$ into $\bm{v}_1, \bm{v}_2$ and ask for $\{\bm{v}_1:\forall \bm{v}_2 \, . \, A(\bm{v}_1,\bm{v}_2)\Rightarrow H(\bm{v}_1,\bm{v}_2)\}$. The result is a formula in the free variables $\bm{v}_1$ that weakens or strengthens the assumptions.  If generated automatically the economist gains information about how to reformulate assumptions that justify her hypothesis. 

\subsection{Quantifier Elimination}

Such problems fall within the framework of \emph{Quantifier Elimination} (QE): the generation of an equivalent quantifier-free formula from one that contains quantifiers.  QE is known to be possible over real closed fields thanks to the seminal work of Tarski \cite{Tarski1948}.  Practical implementations followed with Collins' Cylindrical Algebraic Decomposition \cite{Collins1975} and Weispfenning's Virtual Substitution \cite{Weispfenning1988}.  There are modern implementations of QE in many computer algebra systems.
% \textsc{Mathematica} \cite{Strzebonski2006}, \textsc{Redlog} \cite{DS97a}, \textsc{Maple} (\textsc{SyNRAC} \cite{IYA14} and the \textsc{RegularChains}  Library \cite{CM14c}) and \textsc{Qepcad-B} \cite{Brown2003b}.

QE has found many applications within engineering and the life sciences.  
Recent examples include:  
the derivation of optimal numerical schemes \cite{EH16}, 
%artificial intelligence to pass university entrance exams \cite{WMTA16}, 
weight minimisation for truss design \cite{CC18}, 
and biological network analysis \cite{BDEEGGHKRSW17}.
%, \cite{EEGRSW17}.
%The recent survey article \cite{Sturm2017} has applications in geometric theorem proving, verification and the life sciences.
However, applications in the social sciences are lacking (the nearest we can find is \cite{LW14}).
%, though there is also the thread of work in \cite{CKLR15}.
%}.  
On the few occasions when QE has been mentioned in economics it has been dismissed as infeasible, e.g. %too "computationally demanding" \cite{CDFQ14} and 
``something that is do-able in principle, but not by any computer that you and I are ever likely to see'' \cite{Steinhorn2008}.  But that dismissal is based on theoretical complexity results rather than experience with actual software applied to actual economic reasoning.  Many meaningful economics problems can be studied with modern QE implementations\footnote{A dataset of 45 economic reasoning examples that may be tackled with QE technology is available here: \url{https://doi.org/10.5281/zenodo.1226892}}, with the barrier to further use acceptance by the community, and experience with the software.

\subsection{A New Mathematica Package TheoryGuru}

This motivated the development of a new tool to aid the application of QE to economics: a package called \texttt{TheoryGuru} to run in the \textsc{Mathematica} computer algebra system \cite{Wolfram2000}.  This is able to parse input from economists, run some error and sanity checks, and then utilise \textsc{Mathematica}'s QE tools and offer interpretations of the results.  These QE tools are accessed by the \textsc{Mathematica} \texttt{Resolve} command with some of the underlying algorithms described in \cite{Strzebonski2006}, \cite{Strzebonski2010}, \cite{Strzebonski2012}, \cite{Strzebonski2016}.
The paper proceeds in Section \ref{SEC:Functionality} by introducing the functionality of \texttt{TheoryGuru}.  Then in the remaining sections we describe examples of its use.

%------------------------------------------------------------
\section{Functionality}
\label{SEC:Functionality}

\subsection{Main Functionality}

The purpose of \verb+TheoryGuru+ is to lower the costs of applying QE to economics.  Hence it assumes the expression of reasoning in the format traditional to the field: as a conclusion to be possibly deduced from a set of assumptions. 

\noindent 
The core functionality of \texttt{TheoryGuru} is then as follows:
\begin{description}
\item[Check for errors:] Provide warnings on likely typographical errors in variables (e.g., when a variable appears only once in the entire formula) or formula structure (e.g. the user may have confused \texttt{=} with \texttt{==}).
\item[Parse input:]  This includes identifying from context whether a variable is a vector, scalar, or boolean;  processing input given in a \emph{pretty} mathematical notation (e.g. derivatives) into a format accepted by \texttt{Resolve}; standardizing dot products and integrals (e.g., distribute plus and alphabetically sort arguments of commutative operators).
\item[Adding standard assumptions:] If dot products are present, then add to user assumptions the necessary and sufficient conditions for the Gramian matrix (representing dot products for all pairs of vectors) to be positive semi-definite.  This rules out vectors with imaginary elements.
\item[Check assumptions:]  The package will next check that the assumptions provided are not mutually contradictory: the situation of the bottom left entry in Table \ref{tab:main}.  This is done via a call to \texttt{Resolve} to check there is at least one solution to $A(\bm{v})$ $-$ a fully existentially quantified QE sub-problem.  
\item[Form main QE input:] Automatically assemble the two Tarski formulae for the main calls (as given in Table \ref{tab:main}).
\item[Make algorithm choices:] Currently this refers to (a) whether to process a universal or existential sentence and (b) the variable ordering determining the sequence for eliminating quantifiers.  It is well known that the choice (b), while not affecting the correctness of the output, can have a large effect on computational resources required \cite{HEWDPB14}.
\item[Output interpretation:] Then after making the two calls to \textsc{Mathematica}'s \verb+Resolve+, the package interprets the results by identifying the relevant cell from Table \ref{tab:main}.  The package also suggests what to do next: e.g., when applicable, show a counterexample, solve simultaneous equations appearing in the assumptions, or redo the QE with some free variables.
\end{description}

\subsection{Access and Documentation}

\begin{figure}
	\centering
	\includegraphics[width=0.5\linewidth]{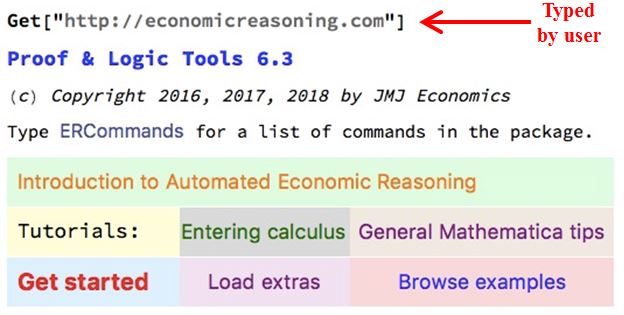}
	\caption{Initial load screen of \texttt{TheoryGuru}}
	\label{fig:screenshot01loadpackage}
\end{figure}

To access \texttt{TheoryGuru} the reader will need a modern version of \textsc{Mathematica}\footnote{The \texttt{Resolve} function has evolved and improved over the years and so the performance of \texttt{TheoryGuru} will alter correspondingly.} and then installation follows from simply running the command: \\
\verb+Get["http://economicreasoning.com"]+
which produces an interface as in Figure \ref{fig:screenshot01loadpackage}.  Not only does this install the underlying code, it also provides links to tutorials, tips, help and a large bank of examples (as shown in Figure \ref{fig:screenshot01loadpackage}). The examples are also available online at \url{http://examples.economicreasoning.com/} as both interactive \textsc{Mathematica} notebooks and static pdfs.

The main functionality is accessed via the function \texttt{TheoryGuru} which requires two arguments: a collection of assumptions and a hypothesis.

%------------------------------------------------------------
\section{Examples of TheoryGuru Use}
\label{SEC:Examples}

\subsection{Tax Incidence}

\begin{figure}
	\centering
	\includegraphics[width=0.8\linewidth]{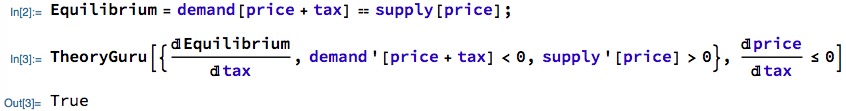}
	\caption{Tax Incidence example in \texttt{TheoryGuru}}
	\label{fig:screenshot02firstqe}
\end{figure}

Our first, admittedly simple, \emph{Tax Incidence} example is about the effect of a tax on buyers and sellers in a market.  Each transaction involves the buyer paying \texttt{price} to the seller in addition to paying \texttt{tax} to the government.  The symbolic functions \texttt{demand(.)} and \texttt{supply(.)} represent the quantities that buyers purchase and sellers sell, respectively, as a function of the price they pay or receive (so for the buyer that includes the tax).  A market equilibrium \texttt{price} has the quantity demanded equal to the quantity supplied.  The equilibrium condition can be input to \textsc{Mathematica} as shown in the top cell of Figure \ref{fig:screenshot02firstqe}, which assigns the condition the natural language name \verb+Equilibrium+.

The first argument of the call to \texttt{TheoryGuru} in the second cell of Figure \ref{fig:screenshot02firstqe} is a set of assumptions.  The first of these is that changing the \texttt{tax} changes the market from one equilibrium to another\footnote{The notation is consistent with an economist saying that she ``totally differentiates the equilibrium condition''.  This differentiation is automatically performed by \textsc{Mathematica} when the \texttt{TheoryGuru} function is evaluated.}.  The remaining two constrain the slope of the demand and supply curves in the neighborhood of the market equilibrium.  The second argument is the hypothesis the user wishes to test, in this case whether the price impact is negative or zero.

The call causes \texttt{TheoryGuru} to automatically assemble Tarski formulae,  
%for the assumptions and for the hypothesis, 
in which it recognizes \texttt{demand}\textquotesingle\texttt{(price+tax)} and \texttt{supply}\textquotesingle\texttt{(price)} as partially interpreted functions \cite[page 73]{KS13b}.  Following the generic format presented above, there are two QE problems for \texttt{TheoryGuru} to consider: the existence of an example and the existence of a counterexample.  Here the output is simply \texttt{True} because there is no counterexample: i.e., no way to have a positive price impact while satisfying the assumptions.

\begin{figure}
	\centering
	\includegraphics[width=0.6\linewidth]{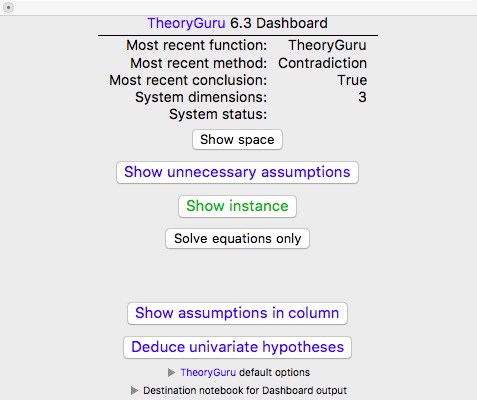}
	\caption{The \texttt{TheoryGuru} dashboard}
	\label{fig:screenshot03dashboard}
\end{figure}

\begin{figure}
	\centering
	\includegraphics[width=0.8\linewidth]{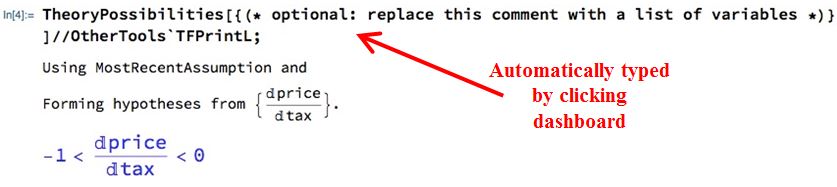}
	\caption{\texttt{TheoryPossibilities} call from dashboard button to propose new hypothesis}
	\label{fig:screenshot04theorypossibilities}
\end{figure}

When \texttt{TheoryGuru} evaluates, a dashboard (Figure \ref{fig:screenshot03dashboard}) appears summarizing the calculation and offering the user possible next steps.
In the tax incidence example, the user may be wondering what else can be concluded about the price impact.  The button labelled \textquotesingle\textquotesingle
Deduce univariate hypotheses\textquotesingle\textquotesingle \, on the dashboard serves this purpose.  Pressing it automatically generates a call to the function \texttt{TheoryPossibilities} as shown in Figure \ref{fig:screenshot04theorypossibilities}.  Here, one or more free variables are provided by the user, or else variables are chosen by the software (giving priority to total derivative variables and alphabetical order).  The assumptions are then projected on each of the free variables separately (eliminating existential quantifiers from all variables except that one), with the resulting formulae simplified.  In this example we discover the price impact must be strictly between $-1$ and $0$. 

\begin{figure}
	\centering
	\includegraphics[width=0.7\linewidth]{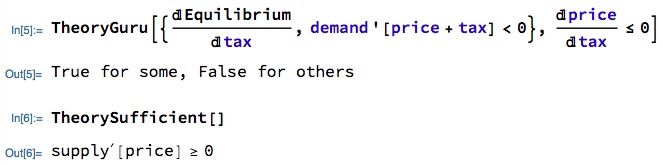}
	\caption{Example use of \texttt{TheorySufficient} to recommend additional assumptions}
	\label{fig:screenshot05theorysufficient}
\end{figure}

Users can be forgetful or have an imperfect understanding of an economic model.  In the top cell of Figure \ref{fig:screenshot05theorysufficient} no definitive conclusion about price impact is reached because the user has forgotten to constrain the supply curve's slope.

The forgotten assumption can be discovered with \verb+TheorySufficient+.  It assembles the formula $A \land \lnot H$ defining counterexamples.  It then projects that set on each of the axes (three in this example).  It then shows the disjunction of each formula, after each is simplified based on the assumptions and then negated.  Here two formulae are discarded because they are \verb+False+ or identical to $H$.  The third is the missing supply-slope restriction output in the second cell of Figure \ref{fig:screenshot05theorysufficient}.  Note that, by construction, any of \verb+TheorySufficient+'s disjunction branches, together with the user's (insufficient) assumptions, imply the user's hypothesis.

\subsection{Gender Wage Gap}

\begin{figure}[h]
	\centering
	\includegraphics[width=0.6\linewidth]{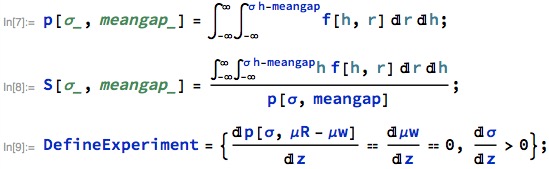}
	\caption{Defining variables for the Gender Wage Gap example}
	\label{fig:screenshot06adoubleintegralinput}
\end{figure}

We now look at a more involved \emph{Gender Wage Gap} example that studies the effect of wage inequality on women's supply of human capital to the market.  Women are assumed to have (possibly correlated) skills $h$ and $r$ in market work and non-market activities, respectively.  These skills have a population distribution modelled with the joint density function $f(h,r)$, which is normalized to have unconditional means of zero.  Women work if and only if their non-market log wage $r + \mu_r$ is less than $\sigma h + \mu_w$, their market log wage
%\footnote{It follows that mean non-market and market log wages are $\mu_r$ and $\mu_w$, respectively.}.
.  It follows that mean non-market and market log wages are $\mu_r$ and $\mu_w$, respectively.
The employment rate of women is $p(\sigma,\mu_w-\mu_r)$ as defined in the top cell of Figure \ref{fig:screenshot06adoubleintegralinput}, and the average skill in the market is $S(\sigma,\mu_w-\mu_r)$ as defined in the next cell.  

\begin{figure}[t]
	\centering
	\includegraphics[width=0.99\linewidth]{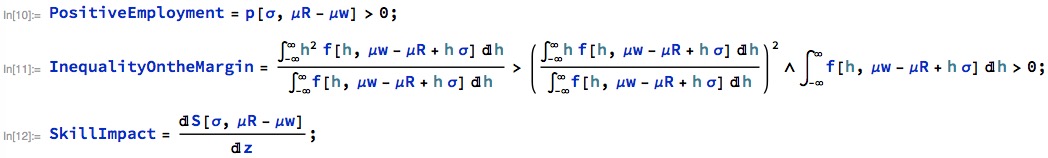}
	\caption{Further definitions and restrictions to the Gender Wage Gap example.}
	\label{fig:screenshot06bdoubleintegralinput}
\end{figure}

\begin{figure}[t]
	\centering
	\includegraphics[width=0.99\linewidth]{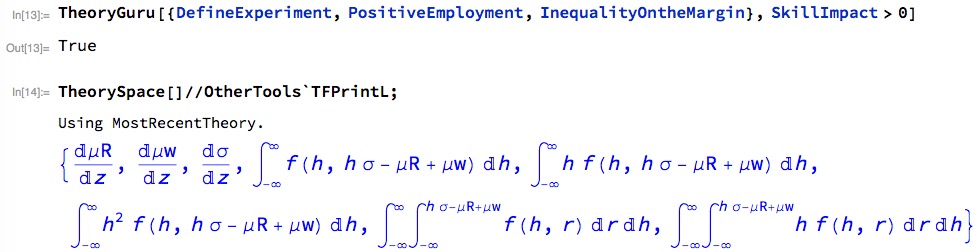}
	\caption{Evaluating the Gender Wage Gap example}
	\label{fig:screenshot07doubleintegralqe}
\end{figure}

In \cite{MR08} the Gaussian model is used to show how wage inequality, as modelled by $\sigma$, affects the average skill in the market.  However, the \emph{selection rule} result -- that is, the effect of $\sigma$ on $S$ holding constant $p$ by varying $\mu_r$ -- does not require the Gaussian assumption.
To show this, we define $z$ to be any change in $\sigma$ and $\mu_r$ that increases $\sigma$ and holds $p$ constant, as defined in the third cell of Figure \ref{fig:screenshot06adoubleintegralinput} (\verb|DefineExperiment|).  Figure \ref{fig:screenshot06bdoubleintegralinput} then assigns natural language to two definitions (top and bottom cells) as well as restrictions on partially interpreted functions (middle cell).
The top cell of Figure \ref{fig:screenshot07doubleintegralqe} shows that a positive skill impact can be deduced from the assumed properties of the partially interpreted functions.  In economics terms: inequality increases the average skill that women supply to the market, thereby narrowing the measured wage gap with men.

At first glance, the gender wage gap example appears to involve integrable probability density functions rather than the scalar variables required by the QE algorithms employed by \textsc{Mathematica}'s \verb|Resolve| function.  But the reasoning in this and many other examples depends on the probability density functions only as they are summarized by various scalars.  \verb|TheoryGuru| automatically discovers these scalar variables, which can be viewed by the user who clicks \textquotesingle\textquotesingle
Show space\textquotesingle\textquotesingle \, on \verb|TheoryGuru|'s dashboard.  The result of that click is the last cell of Figure  \ref{fig:screenshot07doubleintegralqe}.

\section{Run Times}
\label{SEC:Timings}

\begin{figure}[t]
	\centering
	\includegraphics[width=0.9\linewidth]{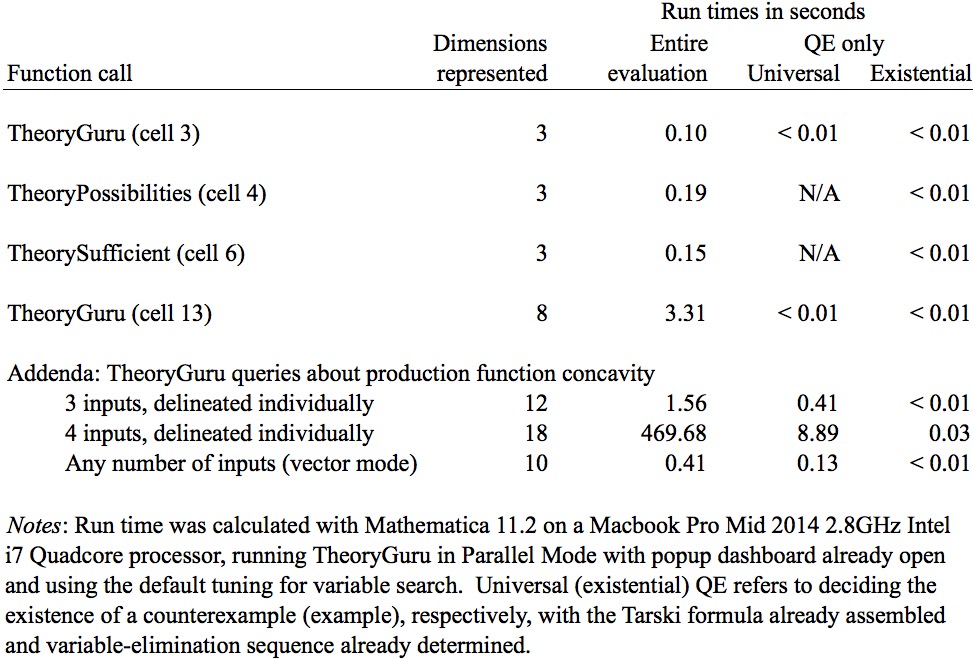}
	\caption{Run times for several examples}
	\label{fig:screenshot08runtimes}
\end{figure}

Figure \ref{fig:screenshot08runtimes} shows the run times for several of the function calls shown above.  As explained, each evaluation involves preparation of a QE problem for \textsc{Mathematica}'s \texttt{Resolve} function, followed by that QE call.  The cell numbers refer to those used by \textsc{Mathematica} in the screen shots above.
The figure's first run time column is for the entire evaluation of the \texttt{TheoryGuru} command.  The next column shows, when applicable, the amount of time it took for just the ``universal'' QE regarding the existence of a counterexample.  The final column is the amount of time it took for just the ``existential'' QE regarding the existence of an example (the faster QE for calls that have no counterexamples).

In order to explore the limits of the software, we consider queries regarding the concavity of quasiconcave production functions, whose economics and algebra we discuss in \cite{MBDET18}.  The three-input version evaluates in less than two seconds.  The four-input version is considerably more complicated, and evaluates in about eight minutes primarily because of a long search to find a relatively efficient order for eliminating quantifiers\footnote{It is once that order is obtained that the corresponding QE needs only 8.89 seconds.}.  The problem can be solved more elegantly with vectors, with a quicker run time as shown in the final row (see also \cite{nber16}).

%------------------------------------------------------------
\section{Final Thoughts}
\label{SEC:Final}

We have demonstrated how economic reasoning may be automated using QE procedures and how the \texttt{TheoryGuru} tools greatly reduce the costs to an economist of accessing that technology.  We note that a set of benchmark examples that originate from economics and may be tackled with QE is now available at \url{https://doi.
org/10.5281/zenodo.1226892} and described in \cite{MBDET18}.  Future work will involve considering how the underlying QE technology could be optimised for such examples, whose structure is often not well represented in the broader QE literature.

%------------------------------------------------------------

\bibliographystyle{splncs03}
\bibliography{CAD}

\end{document}